\documentstyle[12pt]{article}
\newcommand{\NP}[1]{Nucl. \ Phys.}
\newcommand{\PL}[1]{Phys. \ Lett.}
\newcommand{\p}[1]{\partial}
\newcommand{\PRL}[1]{Phys.\ Rev.\ Lett. }
\newcommand{\AP}[1]{Ann.\ Phys. }
\newcommand{\IM}[1]{Inv.\ Math. }
\newcommand{\JMP}[1]{ J.\ Math.\ Phys. }
\newcommand{\RMP}[1]{Rev.\ Mod.\ Phys.}
\newcommand{\JDG}[1]{ J.\ Diff.\ Geom. }
\newcommand{\PTP}[1]{ Prog.\ Theor.\ Phys. }
\newcommand{\SPTP}[1]{Suppl.\ Prog.\ Theor.\ Phys. }
\newcommand{\PR}[1]{Phys.\ Rev. }
\newcommand{\PREP }[1]{Phys.\ Reports }
\newcommand{\NC}[1]{Nuovo \ Cim. }
\newcommand{\NCL }[1]{Nuovo\ Cim.\ Lett. }
\newcommand{\CMP}[1]{Commun.\ Math.\ Phys. }
\newcommand{\TMF}[1]{Theor.\ Math.\ Phys. }
\newcommand{\CQG}[1]{Class.\ Quant.\ Grav. }
\newcommand{\FAA}[1]{Funct. Analys. Appl. }
\newcommand{\JP}[1]{J.\ Phys. }
\newcommand{\JA }[1]{ J. Algebra }
\newcommand{\JFA}[1] {J. Funct. Anal. }
\newcommand{\MPL}[1] { Mod. Phys. Lett. }
\newcommand{\IJMP}[1] { Int. J. Mod. Phys. }
\newcommand{\RMAP}[1] { Rep.\ Math.\ Phys. }
\begin{document}
\title{\begin{flushright}
{\small Preprint SMI-12-93\\ June, 1993}
\end{flushright}
\vspace{2cm}
$D=2$ string theory in
target space/world-sheet light-cone gauge.}
\author{A.P.Zubarev
\thanks{Supported in part by RFFR under grant N93-011-147}
\\Steklov Mathematical Institute\\
Vavilov st. 42, Moscow, 117966, GSP-1, Russia}

\maketitle
\begin{abstract}
$D=2$ free string in linear dilaton background is considered
in  so called
target space/world-sheet light cone gauge
$ X^{+}=0,~g_{++}=0,~g_{+-}=1$.
After gauge fixing the theory has the residual
Virasoro and  $U(1)$ current symmetries.
 The physical spectrum related to $SL_2$
invariant vacuum is found to be trivial.
We find that the theory has a nontrivial spectrum if
 the states in different non-equivalent representations
 ("pictures") of CFT algebra
of matter fields are considered.
$$~$$
\end{abstract}

%%%%%%%%%%%%%%%%%%%%%%%%%%%%%%%%%%%%%%%%%%%%%%%%%%%*****************
\def\d{{\cal D}}

%*********************************************************************
$$~$$
$$~$$

It has been advocated for a long time that a string theory is a prime
candidate for a unified quantum description of high energy physics.
However, in present our understanding of string theory is far from
being complete. Among the unsolved problems the main one is to find a
non-trivial string vacuum related to real physics. The investigation of
these questions is hindered by complexity of string theory. In this way
it is very instructive to deal with a toy model which posses
a lot  of important features of conventional $D=26$ critical
string theory but which is more simple for complete description.
The string theory in two
dimensional target space ($D =2$ string theory)
is an
appropriate candidate for such a toy model.

During last years the $D=2$ string theory was the subject of intense study.
The understanding of this theory has been greatly improved by resent
progress in matrix models \cite{0}. However, it is very important
to understand the results of matrix models description of the theory
in the context of the usual continuous approach.
It is well-known that $D=2$ string appears as an effective theory
after quantization of 2d gravity coupled to $c=1$ matter
 by the action
\begin{equation}
\label{a1}
S={1\over 8\pi}\int\sqrt{g}g^{\alpha\beta}\partial_\alpha
x\partial_\beta x
\end{equation}
\noindent
In the conformal gauge
$g_{\alpha\beta}=e^{2\phi}\hat g_{\alpha\beta}$ the quantization of
the action (1) is reduced to
the quantization of the conformal factor $\phi$ of the
metric with the Liouville action \cite{1}. Originally the
problem of quantization  of the action (\ref{a1})
 was solved by Polyakov in light-cone gauge
\cite{2}. He had quaintized the theory by solving the anomalous Ward
identities in terms of SL(2,R) Kac-Moody algebra. The quantization of
the theory in the conformal gauge  was obstructed by
complexity of quantum measure of the Liouville field $\phi$. Distler and
Kavai \cite{3} were the first who appreciated that the quantum measure
over $\phi$ is not the standard translation invariant quantum field
theory measure. They had proposed to replace the non-standard measure
by a translation invariant one times a Jacobian factor which appears due the
changing of the measure. This Jacobian was assumed to be  of the
form of the exponential of the original Liouville action with
renormalized background charge and coupling constant.  The
same result was obtained independently by David \cite{4}. Later it was
shown by Mavromatos and Miramontes \cite{5} and D'Hoker and Kurzepa
\cite{6} that conjecture of \cite{3,4} can be derived in a more direct way
by regularization the formal expression for the Jacobian as a functional
determinant of a rather singular operator using the heat-kernel method.
The final expression for partition function of the theory in conformal
gauge contains an integrations over $c=1$ matter field $x$ and
Liouville field $\phi$ with the translation invariant and $\phi$-independent
measure and may be considered as  the conformal invariant theory in $2d$
target space with coordinates $X_\mu =(x,\phi )$.
Another way to obtain this result is to start  from the string
action in $2d$ target space  in linear dilation background
\begin{equation}
\label{a2}
S(X_\mu ,g_{\alpha\beta})
=\frac{1}{8\pi}\int\sqrt{g}(g^{\alpha\beta}\partial_\alpha
X_\mu \partial_\beta X^\mu +Q_\mu X^\mu R)
\end{equation}
Classically the action
(\ref{a2}) is invariant under world-sheet reparametrization but is
not invariant
under Weyl rescaling because of the curvature term. However the Weyl invariance
is recombined after quantization for the "critical" values of the background
charge $Q^2=8$. Then the theory becames the critical one and one must divide
the path integral measure by  the Weyl group volume. By choosing a
conformal gauge one obtains the form of path integral partition
function that was conjectured in \cite{3,4}.

One of the important features of the  model is a very
interesting spectrum of the physical states. It consist of a lowest
continuous excitation ("tachyon") plus an infinite series of the states
with special fixed values of energy and momentum (discrete
states). This fact was firstly recognized in the matrix model
description \cite{7}. The appearance of discrete states seems rather
transparent if the model is described as $D=2$ string
\cite{8,9}. Since we deal with a string one can expect an infinite
number of higher level excitations. But since the theory is
2-dimensional gauge invariant theory one can also expect that all these
states can be eliminated by a gauge transformations. However
 not all of states are pure gauge due to the nontrivial
background. The states
which can not be gauge away can be understood as remnants of
transversel string excitations.

Last time the studies of discrete states have been the subject of a
number of papers \cite{10} - \cite{17}.
It was understood that these states can be
treated as non-trivial cohomology classes of the corresponding BRST
charge (Lian and Zuckerman \cite{10}). The non-trivial cohomology was
found with ghost numbers 0,...,3 (Bouwgknegt, McCarthy and Pilch
\cite{11}) and it was discovered ( Klebanov and Polyakov \cite{12},
Witten \cite{13}, Witten and Zwiebach \cite{14}) that these states forms
the remarkable algebraic structure. Namely the spin zero ghost number
zero discrete states generate a ground ring and the spin one ghost
number one discrete states generate $W_\infty$ current algebra, which
is the algebra of the symmetry of the theory. This symmetry have been
used to facilitate the calculation of tachyon amplitudes (Klebanov
\cite{15}) and has also been uncovered in matrix models \cite{7}.

It is necessary to note that the discussion around a discrete states
and their role in the theory was mainly related to  the
conformal gauge. In the recent work of Marcus and Oz \cite{16} the physical
spectrum of the theory was studied in the Polyakov light-cone gauge. It
was found that if the state space of the gravitational sector of the
theory is taken to be the irreducible representations of the SL(2,R)
current algebra. Then the physical spectrum of the theory is not the same as
that in the conformal gauge. However
the physical spectrum of the
theory in light cone gauge is the same as in the theory in conformal gauge
in the different Wakitomo free field
representation of gravitational field.

A different light-cone gauge was used in the work of Smith \cite{17}.
He considered the theory with the action (\ref{a2}) by imposing the gauge
conditions on target space light-cone coordinates
$X^+=\zeta ^0$
as well as on the world-sheet metric $g_{--}=0,~\sqrt{-g}=1$.
As was concluded in \cite{17} the theory contains only two-dimensional
tachyon and no  exited states.

 From the general reasons
the physical content of the theory
such as the physical spectrum should  be expected to be
independent on the
gauge choice. So it is interesting to understand the relation of the
theory in the conformal gauge to the theory in other gauges.

In this work we study the $D=2$ string in a special
gauge which we call "the target
space/world-sheet light cone" gauge. We shall
quantize the theory by using the  path integral.
As it will be shown after the
gauge fixing the theory has the residual Virasoro and  U(1) current
symmetries. We calculated a conserved currents associated with these
symmetries. It  occurs that the structure of the currents
is the same as was obtained in \cite{17} for other gauge choice.
The matter sector of the theory has the central charge $c=28$ which
is canceled by the central charge of ghost system $c^{gh}=-28$.
The corresponding BRST charge  $Q$ in nilpotent $Q^2=0$.
Also we shall discussed  a physical
spectrum of the theory. We find that it depends on the choice of
representations of states of the theory. In particular the physical space built
out SL$_2$ invariant vacuum is trivial. We find some of the physical states
lying in  different non-equivalent representations ("pictures") of CFT
algebra of the fields.

We start with $D=2$ string action in a linear dilation background in the form
of (\ref{a2}):  \begin{equation} \label{a3} S(X,g)={1\over 8\pi}\int\,\,
d^2\zeta\sqrt{-g}(g^{\alpha\beta}\partial_\alpha X_\mu D_\beta X_\nu
+Q_\mu X_\nu R)\eta^{\mu\nu}.
\end{equation}
Here the integration is performed over world-sheet with topology of
sphere, $\zeta^\alpha =(\zeta^1,\zeta^2)$ are a world-sheet
coordinates, $g_{\alpha\beta}$ is a world-sheet metric of
pseudoEuclidean signature, $R$ is a Ricci scalar, $X_\mu =(x,\phi )$ are
target  space coordinates, $\eta_{\mu\nu}=diag (+,+)$ is a flat
target space metric and $Q_\mu =(0,2\sqrt{2})$ is a background charge.

The partition function of the theory
\begin{equation}
\label{a4}
Z=\int\,\,\d g_{\alpha\beta}\d X_\mu e^{iS(x,g)}
\end{equation}
is invariant under an arbitrary combination of the world-sheet
diffeomorphisms and Weyl rescaling
$$\delta X_\mu =\partial_\alpha
X_\mu \varepsilon^\alpha -Q_\mu\varepsilon,$$
\begin{equation}
\label{a5}
\delta g_{\alpha\beta}=\partial_\gamma
g_{\alpha\beta}\varepsilon^\gamma +g_{\alpha\gamma}
\partial_\beta\varepsilon^\gamma +g_{\beta\gamma}\partial_\alpha
\varepsilon^\gamma .
\end{equation}
Here $\varepsilon^\alpha$ is an infinitesimal vector field generating a
diffeomorphism and $\varepsilon$ is an infinitesimal Weyl parameter.
Note that while the diffeomorphism invariance is hold on both classical
and quantum levels, the Weyl transformation are only the symmetry of
quantum theory.

Following the standard prescriptions we have to fix the invariance (\ref{a5})
by imposing an appropriate gauge conditions. We choose combined target
space/world-sheet gauge conditions which are
\begin{equation}
\label{a6}
X^+=0\quad ,\quad g_{+-}=1\quad ,\quad g_{--}=0.
\end{equation}
Here $X^\pm ={1\over\sqrt{2}}(X^1\pm iX^2)$ are light-cone
coordinates of target space and $g_{+-},~g_{--}$ are the components of
the metric $g_{\alpha\beta}$ in light-cone world-sheet coordinate
basis $\zeta^\pm =\zeta^1\pm\zeta^2$.

In the gauge (\ref{a6}) the gauge transformations (\ref{a5}) read
\begin{eqnarray}
\delta
X^-&=&\partial_-X^-\varepsilon^-+\partial_+X^-\varepsilon^+-Q^-\varepsilon ,
\nonumber\\
\delta X^+&=&-Q^+\varepsilon ,\nonumber\\
\delta g_{++}&=&(2\partial_++\partial_-h)\varepsilon^-+(\partial_+h+
2h\partial_+)\varepsilon^+ +2h\varepsilon ,\nonumber\\
\delta
g_{+-}&=&\partial_-\varepsilon^-+
(\partial_++h\partial_-)\varepsilon^++2\varepsilon ,
\nonumber\\
\delta g_{--}&=&2\partial_-\varepsilon^-
\label{a7}
\end{eqnarray}
(here and below we use the notation $g_{++}=h$).

The gauge conditions (\ref{a6}) do not fix the symmetry (\ref{a5})
completely. Indeed,
the eqs. (\ref{a6}) remain unchanged under the transformations (\ref{a7})
with parameters
\begin{equation}
\label{a8}
\varepsilon^+=v(\zeta^+)\,\, ,\,\,
\varepsilon^-=u(\zeta^+)-\zeta^-\partial_+ v(\zeta^+)\,\, ,\,\,
\varepsilon =0.
\end{equation}
In terms of $v$ and $u$ the transformations of $X^-$ and $h$ are
\begin{eqnarray}
\delta
X^-=\partial_-X^-u(\zeta^+)+(\partial_+X^--
\zeta^-\partial_-X^-\partial_+)v(\zeta^+),
\nonumber\\
\delta h=(2\partial_++\partial_-h)u(\zeta^+)
+(\partial_+h+2h\partial_+-2\zeta^-\partial_+^2
-\zeta^-\partial_-h\partial_+) v(\zeta^+) .
\label{a9}
\end{eqnarray}
One may easely verify that the transformations (9) are the symmetries of
gauge fixed action
\begin{equation}
\label{a10}
S(X^-,h)={Q^+\over 8\pi}\int\,\, d^2\zeta h\partial_-^2X^-.
\end{equation}
The partition function (\ref{a4}) in the gauge (\ref{a6}) takes the form
\begin{equation}
\label{a11}
Z=\int\,\, \d h\d X^-\Delta^{FP}e^{iS(X^-,h)}
\end{equation}
where $\Delta^{FP}$ is the Faddeev-Popov determinant which can be
written in terms of path integral over ghosts $(c^\pm ,c)$ and
antighosts $(b_{++},b_{+-},b)$ with the ghost action
\begin{eqnarray}
S_{gh}=\int d^2\zeta\sqrt{-g}[{1\over
2}g^{\alpha\beta}g^{\gamma\delta} b_{\alpha\gamma}\delta
g_{\beta\delta}+2b \delta X^+]_{\varepsilon^\alpha = c^\alpha
,\varepsilon = c,b_{--}=0}=\nonumber\\
=\int d^2\zeta
[b_{++}\partial_-c^++b_{+-}\partial_-c^-+b_{+-}\partial_+c^+- b_{+-}h
\partial_-c^++2b_{+-}c-2Q^+bc].
\label{a12}
\end{eqnarray}

The antighost $b$ in (\ref{a12}) plays the role of Lagrange multiplier and
integration over it simply produces a $\delta$-functional which
constraints the value of ghost $c$ to be zero. Then integrating over
$c$ one obtains the ghost action in the form (\ref{a12}) with $b=c=0$. Next by
changing
\begin{eqnarray}
c^-&=&\eta -\zeta^-\partial_+c , ~~ c^+~=~c,\nonumber\\
b_{+-}&=&b+(h-\zeta^-\partial_+)\xi,~~  b_{+-}~=~\xi
\label{a13}
\end{eqnarray}
the ghost action  casts into the form
\begin{equation}
\label{a14}
S_{gh}=\int d^2\zeta [b\partial_-c+\xi\partial_-\eta ].
\end{equation}
The residual transformations for ghosts $bc$ and $\xi\eta$  are written by
taking into account their transformation properties
\begin{eqnarray}
\delta c&=&(-c\partial_-+\partial c)v(\zeta^+),\nonumber\\
\delta b&=&-\partial_-\xi
u(\zeta^+)+(2b\partial_-+\partial_-b)v(\zeta^+),\nonumber \\
\delta\eta &=&-(c\partial_-+\partial_-c)u(\zeta^+)+(\eta\partial_-
+\partial_-\eta )v(\zeta^+),\nonumber\\
\delta\xi &=&\partial_-\xi v(\zeta^+)
\label{a15}
\end{eqnarray}
and leave the ghost action (\ref{a14}) unchanged.

Presence of the residual symmetries implies an existence of conserved
currents. These currents can be found by varying the matter and ghost
actions with general $u=u(\zeta^+,\zeta^-), v=v(\zeta^+,\zeta^-)$:
\begin{eqnarray}
\delta S_M={Q^+\over 8\pi}\int\,\, d^2\zeta
(\partial_-Ju+\partial_-Tv), \nonumber\\
\delta S_{gh}=\int\,\, d^2\zeta (\partial_-J^{gh}u+\partial_-T^{gh}v).
\label{a16}
\end{eqnarray}
After some calculations one obtains
\begin{eqnarray}
T&=&-2\partial_+\partial_-X^-h-\partial_-X^-\partial_+h+
\partial_-h\partial_+ X^-+2\partial_+^2X^-\nonumber\\
&&+(-2\partial_+^2\partial_-X^-+\partial_+\partial_-h\partial_-X^-+
\partial_+\partial_-X\partial_-h)\zeta^-,\nonumber\\
J&=&\partial_-X\partial_-h-2\partial_+\partial_-X,
\nonumber\\
T^{gh}&=&\partial_+\xi\eta -2b\partial_+c-\partial_+bc,\nonumber\\
J^{gh}&=&c\partial_+\xi .
\label{a17}
\end{eqnarray}
By inserting into (\ref{a17}) the solutions of the equations of motion for
$X^-$ and $h$
\begin{eqnarray}
X^-(\zeta )&=&\varphi (\zeta^+)+\zeta^-\gamma (\zeta^+),\nonumber\\
h(\zeta )&=&\beta (\zeta^+)+\zeta^-\chi(\zeta^+)
\label{a19}
\end{eqnarray}
we obtain
\begin{eqnarray}
T&=&\chi \partial_+\varphi +2\partial_+^2\varphi -2\partial_+\gamma\beta
-\gamma\partial_+\beta ,\nonumber\\
J&=&\gamma \chi -2\partial_+\gamma .
\label{a20}
\end{eqnarray}
Now let us determine a correlation functions of the theory. The
simplest way to do this is to use the Ward identities associated with
the residual symmetry. The transformations of the fields $\varphi
,\gamma ,\beta$ and $\chi$ are
\begin{eqnarray}
\delta\varphi &=&\gamma u+\partial_+\varphi v,\nonumber\\
\delta\gamma &=&(\partial_+\gamma -\gamma\partial_+)v,\nonumber\\
\delta\beta &=&(2\partial_++\chi)u+(\partial_+\beta
+2\beta\partial_+)v,\nonumber\\
\delta \chi &=&(\chi \partial_+-2\partial_++\partial_+\chi)v.
\label{a21}
\end{eqnarray}

At this point it is convenient to make a Wick rotation $\zeta^1\to
-i\zeta^1$ and go to complex coordinate $z=\zeta^1 +i\zeta^2$. The
Ward identities  written in terms of OPE looks like
\begin{eqnarray}
T(z)\varphi (w)&\sim&{\partial\varphi (w)\over z-w}\,\, ,\,\,
{}~~~~~~~~~~~~~~~J(z)\varphi (w)\sim {\gamma (w)\over z-w},\nonumber\\
T(z)\gamma (w)&\sim&{-\gamma (w)\over (z-w)^2}+{\partial\gamma (w)\over
z-w}\,\, ,\,\, J(z)\gamma (w)\sim O(1),\nonumber\\
T(z)\beta (w)&\sim&{2\beta (w)\over (z-w)^2}+{\partial\beta (w)\over
z-w}\,\, ,\,\, J(z)\beta (w)\sim {2\over (z-w)^2}+{\chi(w)\over z-w},
\nonumber \\
T(z)\chi(w)&\sim&{-4\over (z-w)^3}+{\chi(w)\over (z-w)^2}
+{\partial \chi(w)\over z-w},~~~J(z)\chi (w) \sim O(1).
\label{a22}
\end{eqnarray}
It is follows from (\ref{a22}) and explicit form of $T$ and $J$ that
correlators of the fields have to be
\begin{equation}
\label{a23}
\langle \chi(z)\varphi (w)\rangle = {1\over z-w}\,\, ,\,\,
\langle \gamma (z)\beta (w)\rangle =
{1\over z-w}.
\end{equation}
Similarly for ghost fields we find
\begin{equation}
\label{a24}
\langle \xi (z)\eta (w)\rangle= {1\over z-w}\,\, ,\,\,
\langle b(z)c(w)\rangle= {1\over z-w}.
\end{equation}
The OPE's of currents are
calculated by using (\ref{a23}) and (\ref{a24}):
\begin{eqnarray}
T(z)T(w)&\sim&{14 \over (z-w)^4}+{2T(w)\over (z-w)^2}+{\partial
T(w)\over z-w}\nonumber\\
T(z)J(w)&\sim&{\partial J(w)\over z-w}\nonumber\\
J(z)J(w)&\sim&O(1)
\label{a25}
\end{eqnarray}
for the matter system and
\begin{eqnarray}
T^{gh}(z)T^{gh}(w)&\sim&{-14 \over (z-w)^4}+{2T^{gh}(w)\over
(z-w)^2}+{\partial T^{gh}(w)\over z-w},\nonumber\\
T^{gh}(z)J^{gh}(w)&\sim&{\partial J^{gh}(w)\over z-w},\nonumber\\
J^{gh}(z)J^{gh}(w)&\sim&O(1)
\label{a26}
\end{eqnarray}
for ghost system. The currents $T$ and $J$ generate an analytic
Virasoro and an analytic $U(1)$ current algebras respectively. It is
interesting to note that the current algebra obtained here is the same as for
alternative choice of gauge conditions used in \cite{17}.

The analytic conformal dimensions $\Delta$ of the fields are determined
by their OPE with $T(z)$. We have
\begin{eqnarray}
\Delta (\varphi )&=&0 ,\,\,\, \Delta (\chi)=1 ,\,\,\, \Delta
(\gamma )=-1 ,\,\,\, \Delta (\beta )=2,\nonumber\\
\Delta (\xi )&=&0 ,\,\,\, \Delta (\eta )=1 ,\,\,\, \Delta
(c)=-1 ,\,\,\, \Delta (b)=2,\nonumber\\
\Delta (J)&=&0 ,\,\,\, \Delta (T)=2.
\label{a27}
\end{eqnarray}

 From (\ref{a26}) one sees that the total central charge of analytic Virasoro
algebra for matter plus ghost system vanishes:
\begin{equation}
\label{a28}
c^{tot.}=c^{X^-,h}+c^{gh}=28-28=0.
\end{equation}
The BRST charge associated with the current algebra (\ref{a26}) is
\begin{equation}
\label{a29}
Q_{BRST}=\int \frac{dz}{2 \pi i}[c(T^{\varphi \chi}+T^{\gamma
\beta }+T^
{\xi \eta}+\frac{1}{2}T^{bc})+\eta J](z).
\end{equation}
By direct calculation one can show that $Q^2=0$.
%%%%%%%%%%%%%%%%%%%%%%%%%%%%%%%%%%%%%%%%%%%%%%%%%%%%%%
%%%%%%%%%%%%%%%%%%%%%%%%%%%%%%%%%%%%%%%%%%%%%%%%%

Now what we are going to do is to discuss the physical spectrum
of the theory.
According to the standard prescription, the physical states are
identified with the non-trivial cohomology of the BRST charge
\begin{equation}
 Q|phys\rangle =0,~|phys\rangle \ne Q|\lambda\rangle.
\label{a30}
\end{equation}
If we are restricted
by  the states lying only in the matter sector the constraint
(\ref{a30}) can be written in equivalent form
$$ L_n|phys\rangle =J_n|phys\rangle =0,~n>0, $$
$$ (L_0-1)|phys\rangle =J_0|phys\rangle =0,$$
\begin{equation}
|phys\rangle \ne \sum _{n>O}[L_{-n}|\lambda _n'\rangle
{}~+~J_{-n}|\lambda _n''\rangle ],
\label{a31}
\end{equation}
where $L_n$ and $J_n$ are the coefficients of mode expansion of the
generators $T(z)$ and $J(z).$ To describe the  states
by CFT vertex operators $V(z)$ the  constraints (\ref{a31}) can be
written in terms of OPE
$$T(z)V(w)=\frac{V(w)}{(z-w)^2} + \frac{\partial V(W)}{z-w}
+ O(1), $$
\begin{equation}
J(z)V(w)=O(z-w).
\label{a32}
\end{equation}
It is not difficult to see that there are no local operators
constructed from the conformal fields and their derivatives having
the OPE with $J(z)$ of the form (\ref{a32}). In terms of Fock space
states this means the following. The conformal $SL_2$ invariant
vacuum $|0\rangle $ defined by
\begin{equation}
\phi _n |0\rangle =0,~n>-\Delta (\phi)
\label{a33}
\end{equation}
for each field $\phi (z)$ of dimension $\Delta (\phi)$ is not $U(1)$
invariant one, because it is not annihilated by $K_0$
$$ J_0|0\rangle = (\chi_{-1}\gamma_{1}-2\gamma _{0})|0\rangle \ne 0. $$
It follows then that in the space with $SL_2$-invariant ground state
 there are no nontrivial states which satisfiy  $J_0|phys\rangle
=0$.

To obtain a non-trivial physical spectrum we have to consider an
extended representation of CFT algebra (\ref{a25}). Firstly let us
bosonize the $\gamma \beta $-system according to \cite{20}
\begin{equation}
\gamma = {\tilde \eta} e^{\phi},~~\beta = e^{-\phi}\partial {\tilde \xi}
\label{a34}
\end{equation}
Here $\phi$ is bosonic field such that
$$c^{\phi}=28,$$
$$\langle \phi (z) \phi (w)\rangle = - \ln(z-w),$$
\begin{equation}
T^{\phi}=-\frac{1}{2}\partial \phi \partial \phi
-\frac{3}{2}\partial ^{2}\phi.
\label{a35}
\end{equation}
The fields ${\tilde \xi}$ and ${\tilde \eta}$ are the fermionic ones of
dimensions $0$ and $1$ respectively.
We have $$c^{{\tilde \xi} {\tilde \eta}}=-2,$$
$$\langle {\tilde \xi}(z) {\tilde \eta} (w) \rangle = \frac{1}{z-w},$$
\begin{equation}
T^{{\tilde \xi}{\tilde \eta}}=\partial
{\tilde \xi}{\tilde \eta} .
\label{a36}
\end{equation}

Secondly it is convenient to use the following representation for the
fields $\varphi $ and $\chi $
\begin{equation}
\varphi = \varphi ^- ,~~ \chi= \partial \varphi ^+,
\label{a37}
\end{equation}
where $\varphi ^{\pm} $ are  a new fields with
$$\langle \varphi ^- (z) \varphi ^+ (w) \rangle = \ln(z-w).$$
Defining
\begin{equation}
\varphi ^{\pm}=\frac{1}{\sqrt{2}}(\varphi _1 \pm \varphi _2)
\label{a38}
\end{equation}
we get the system of bosonic fields $\varphi _1$ and $\varphi _2$ with
central charges $c^{\varphi _1}=-23$ and $c^{\varphi _2}=25$ for which
$$\langle \varphi _1(z) \varphi _1(w)\rangle =
-\langle \varphi _2(z) \varphi _2(w)\rangle =
\ln(z-w),$$
$$ T^{\varphi _1}=\frac{1}{2}\partial \varphi _1 \partial \varphi _1 +
\sqrt{2}\partial ^2 \varphi _1, $$
\begin{equation}
 T^{\varphi _2}=-\frac{1}{2}\partial \varphi _2 \partial \varphi _2 -
\sqrt{2}\partial ^2 \varphi _2.
\label{a39}
\end{equation}
The algebra of the fields $\varphi _{1,2}$ is larger than the
$\varphi \chi$-algebra because the second one does not contains zero mode of
$\varphi ^+$ field.

We shall represent an arbitrary state
in the form (with no $\varphi ^+$ -zero mode dependence)
\begin{equation}
V|0\rangle =\partial ^n\varphi ^+...\partial ^m\varphi ^-
...\partial ^{k+1}\xi ...\partial ^r \phi
e^{q\varphi ^-+p\phi}|0\rangle
\label{a40}
\end{equation}
with $n,m,k,l,r=1,2,...$ and integer $p$.
The lowest states (with no oscillator modes ) satisfying the  constraints
(\ref{a32}) are of the form
$$ e^{q\varphi ^--2\phi}$$
with arbitrary $q$ and
$$e^{-4\varphi ^--\phi}.$$

On the next oscillator level we find five kinds of the states
$$
\partial \xi e^{2\varphi ^--3\phi}, ~~
\eta e^{2\varphi ^-},~~
\eta e^{q\varphi ^--3\phi},$$
$$ \partial \varphi ^-e^{q\varphi ^--3\phi},~~
\partial \phi e^{q\varphi ^--2\phi}.$$
It is interesting to give a complete description of all physical states of
the theory in terms of BRST cohomology. A main question is  the
spectrum of
the theory in such gauge is the same as for the theory in the conformal
gauge \cite{11}.  Let us note in this context that not all the states in
the general form (\ref{a40}) are independent due to the existence of
the picture-changing operator
\begin{equation}
P(z)=\{ Q,{\tilde \xi} (z)\}=c\partial {\tilde \xi}(z)-\xi\partial
\varphi ^+e^{\phi}(z)-2\partial \xi e^{\phi}(z).
\label{a41}
\end{equation}
However, the different pictures apparently are not isomorphic
because the picture-changing operator (\ref{a41})
does not have an inverse.
It is seems quite possible that after identifying some of the states
the correct spectrum of $D=2$ string in the conformal gauge will be
reproduced.
We left the investigation of these questions for future.
$$~$$

{\bf ACKNOWLEDGMENT}

I would like to thank the Belgrade Center of Theoretical Physics for its
hospitality while this paper was prepared and
 A.Bogoevi\v{c}, M.Blagoevi\v{c},
B.Sazdovi\v{c}  and P.B.Medvedev for useful discussions.


\vskip2cm
{\small
\begin{thebibliography}{99}
\bibitem{0} \  D.Gross and A.Migdal, {\it Phys. Rev. Lett.},
{\bf 64} (1990) 127; M.Douglas and S.Shenker, {\it Nucl. Phys.},
{\bf B335} (1990) 635;
E.Brezin and V.Kazakov, {\it Phys. Lett.}, {\bf 236B} (1990) 144.
\bibitem{1} \  A. M. Polyakov, {\it Phys. Lett.}, {\bf 103B} (1981) 207.
\bibitem{2} \    A. M. Polyakov, {\it Mod. Phys. Lett.}, {\bf A 2} (1987) 893.
\bibitem{3} \    J. Distler and H. Kawai, {\it Nucl. Phys.} {\bf B 321} (1989)
509.
\bibitem{4} \    F. David, {\it Mod. Phys. Lett.}, {\bf A 3} (1988) 1651.
\bibitem{5} \    N.E.Mavromatos and J.L.Miramontes,
{\it Mod. Phys. Lett.}, {\bf A4} (1989) 1847.
\bibitem{6} \    E.D'Hoker and P.S.Kurzepa, {\it Mod.Phys.Lett.},
{\bf A5} (1990) 1411; E.D'Hoker, {\it Mod. Phys. Lett.}, {\bf A6}
(1991) 745.
\bibitem{7} \    D. Gross, I. Klebanov and M. J. Newman, {\it Nucl. Phys.} {\bf
B350} (1991) 621; D. Gross and I. Klemanov, {\it Nucl. Phys.} {\bf B
352} (1991) 671.
\bibitem{8} \    A. Polyakov, {\it Mod. Phys. Lett.}, {\bf A 6} (1991) 635
\bibitem{9} \    I. Ya. Arefieva and A. P. Zubarev, {\it Mod. Phys. Lett.} {\bf
A 2} (1992) 677.
\bibitem{10} \    B. Lian and G. Zuckerman,
{\it Phys. Lett.} {\bf 266 B} (1991)
21.
\bibitem{11} \    P. Bouwknegt, J. McCarthy and K. Pilch,
{\it Comm. Math. Phys.}
{\bf 145} (1992) 541.
\bibitem{12} \    I. R. Klebanov and A. M. Polyakov
{\it Mod. Phys. Lett.}, {\bf A
6} (1991) 3273.
\bibitem{13} \    E. Witten, {\it Nucl. Phys.} {\bf B 373} (1992) 187.
\bibitem{14} \    E. Witten and B. Zwiebach, {it Nucl. Phys.} {\bf B377} (1992)
55.
\bibitem{15} \    I. Klebanov, {\it Mod. Phys. Lett.}, {\bf A7}
(1992) 723.
\bibitem{16} \    N. Marcus and Y. Oz, {\it "Discrete States of 2D String
theory
in Polyakov's Light-Cone Gauge"}, preprint TAUP-1962-92.
\bibitem{17} \    E. Smith, {\it "Light-Cone Gauge for 1+1 Strings"}, preprint
UTTG-36-91; E. Smith, {\it Trivial Spectrum of Free 1+1 Light-Cone
Strings"}, UTTG-17-92.
%\bibitem{18} \    P.Bouwknegt, J.McCarthy and K.Pilch, {\it "On physical
%%states
%in 2d (topological) gravity"}, preprint CERN-TH.6645/92;
%\bibitem{19} \   P.Bouwknegt, J.McCarthy and K.Pilch, {\it "Semi-infinite
%cohomology in conformal field theory and 2d gravity"}, Preprint
%CERN-TH.6646/92.
\bibitem{20} \   D.Friedan, E.Martinec and S.Shenker, {\it Nucl. Phys.}
{\bf  B271} (1986) 93.
\end{thebibliography}
}
\end{document}